\documentclass[runningheads]{llncs}
\usepackage{multicol}
\usepackage{multirow}
\usepackage{url}
\usepackage{amsmath}
\usepackage{geometry}
\usepackage{array}
\usepackage{tabu}
\usepackage{bm}
\usepackage{graphicx}
\usepackage{subcaption}
\usepackage{makecell}
\usepackage{diagbox}
\usepackage{xcolor}
\begin{document}
\title{MONITOR: A Multimodal Fusion Framework \\to Assess Message Veracity in Social Networks}
\titlerunning{MONITOR: Assessing Message Veracity in Social Networks}
\author{Abderrazek Azri \inst{1} \and C\'ecile Favre\inst{1} \and Nouria Harbi\inst{1} \and J\'er\^ome Darmont\inst{1} \and Camille No\^us\inst{2}}
\authorrunning{Azri et al.}
\institute{Université de Lyon, Lyon 2, UR ERIC \\ 5 avenue Pierre Mend\`es France, F69676 Bron Cedex, France  \and Université de Lyon, Lyon 2, Laboratoire Cogitamus\\
\email{\{a.azri, cecile.favre, nouria.harbi, jerome.darmont\}@univ-lyon2.fr}\\
\email{camille.nous@cogitamus.fr}}
\maketitle              % typeset the header of the contribution
\begin{abstract}
Users of social networks tend to post and share content with little restraint. Hence, rumors and fake news can quickly spread on a huge scale. This may pose a threat to the credibility of social media and can cause serious consequences in real life. Therefore, the task of rumor detection and verification has become extremely important. Assessing the veracity of a social media message (e.g., by fact checkers) involves analyzing the text of the message, its context and any multimedia attachment. This is a very time-consuming task that can be much helped by machine learning. In the literature,  most message veracity verification methods only exploit textual contents and metadata. Very few take both textual and visual contents, and more particularly images, into account. In this paper, we second the hypothesis that exploiting all of the components of a social media post  enhances the accuracy of veracity detection. To further the state of the art, we first propose using a set of advanced image features that are inspired from the field of image quality assessment, which effectively contributes to rumor detection. These metrics are  good indicators for the detection of fake images, even for those generated by advanced techniques like generative adversarial networks (GANs). Then, we introduce the 
Multimodal fusiON framework to assess message veracIty in social neTwORks (MONITOR), which exploits all message features (i.e., text, social context, and image features) by supervised machine learning. Such algorithms provide interpretability and explainability in the decisions taken, which we believe is particularly important in the context of rumor verification.   
Experimental results show that MONITOR can detect rumors with an accuracy of 96\% and 89\% on the MediaEval benchmark and the FakeNewsNet dataset,  respectively. These results are significantly better than those of state-of-the-art machine learning baselines.

\end{abstract}
\section{Introduction}
\label{sec:introduction}

\begin{figure}[htbp]
	\centering
	\begin{subfigure}[b]{0.45\textwidth}
	\centering
	\includegraphics[width=0.55\textwidth]{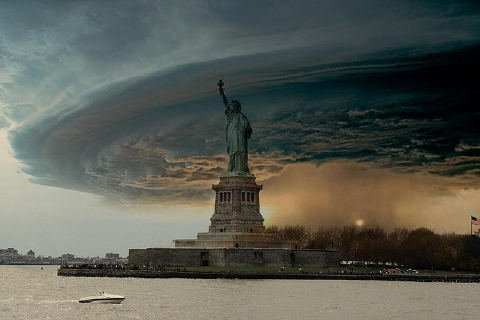}
	\caption{Black clouds in New York City before Sandy!!!}
	\label{}
	\end{subfigure}
	\quad
	\begin{subfigure}[b]{0.45\textwidth}
		\centering
		\includegraphics[width=0.45\textwidth]{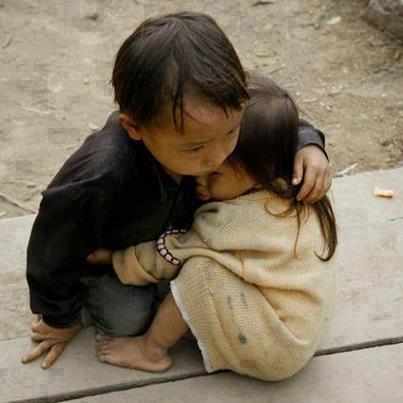}
		\caption{\#NepalEarthquake 4Years old boy protect his little sister. make me feel so sad}
		\label{}
	\end{subfigure}
		\caption{Two sample rumors posted on Twitter}%
		\label{Fig1}%
\end{figure}
After more than two decades of existence, social media has attracted a large number of users. These social platforms allow users to share content and interact with each other. They enable the rapid diffusion of information in real-time, regardless of its credibility, for two main reasons: first, there is a lack of a means to verify the veracity of the content transiting on social media; and second, users often publish messages without verifying the validity and reliability of the information. Consequently, social networks, and particularly microblogging platforms, are a fertile ground for rumors to spread. 

Following previous work \cite{zubiaga2018}, we define a rumor as an item of circulating information whose veracity status is yet to be verified at posting time.
Widespread rumors can pose a threat to the credibility of social media and cause harmful consequences in real life. Thus, the automatic assessment of information credibility on microblogs that we focus on is crucial to provide decision support to, e.g., fact checkers. This task requires to verify the truthfulness of messages related to a particular event and return a binary decision  
stating whether the message is true. 
In the literature, most automatic rumor detection approaches address the task as a classification problem. They extract features from various aspects of messages, which are then used to train a wide range of machine learning \cite{volkova2018misleading} or deep learning \cite{wang2018eann} methods.
Features are generally extracted from the textual content of messages \cite{perez-rosas-etal-2018-automatic} and the social context \cite{wu2018tracing}. However, the multimedia content of messages, particularly images that present a significant set of features, are little exploited.

In this paper, we second the hypothesis that the use of image properties is important in rumor verification. Images may indeed attract more attention than texts \cite{jin2017multimodal}. Furthermore, images play a crucial role in the news diffusion process. For example, in the dataset collected by \cite{jin2017novel}, the average number of messages with an attached image is more than 11 times that of plain text ones. 
Figure~\ref{Fig1} shows two sample rumors posted on Twitter. In Figure~\ref{Fig1}(a), it is hard to assess veracity from the text, but the likely-manipulated image hints at a rumor. In Figure~\ref{Fig1}(b), it is hard to assess veracity from both the text or the image because the image has been taken out of its original context. 
Based on the above observations, we aim to leverage all the modalities of microblog messages for verifying rumors; that is, features extracted from textual and social context content of messages, and up to now unused visual and statistical features derived from images. Then, all types of features must be fused to allow a supervised machine learning classifier to evaluate the credibility of messages.

Our contribution is twofold. First, we propose the use of a set of image features inspired from the field of image quality assessment (IQA) and we prove that they contribute very effectively to the verification of message veracity. These metrics estimate the rate of noise and  quantify the amount of visual degradation of any type in an image. They are proven to be good indicators for detecting fake images, even those generated by advanced techniques such as generative adversarial networks (GANs) \cite{goodfellow2014generative}.  To the best of our knowledge, we are the first to systematically exploit this type of image features to check the veracity of microblog posts. 
Our second contribution is the Multimodal fusiON framework to assess message veracIty in social neTwORks (MONITOR), which exploits all types of message features (i.e., text, social context and image features) by supervised machine learning. This choice is motivated by two factors. First, these techniques provide explainability and interpretability about the decisions taken. We believe that such explanations are necessary, especially in the context of rumors, with people's privacy in line. Second, we do also want to explore the performance of deep machine learning methods in the near future, especially to  study  the  tradeoff between classification accuracy, computing complexity, and explainability.

Eventually, extensive experiments conducted on two real-world datasets demonstrate the effectiveness of our rumor detection approach. 
MONITOR indeed outperforms all state-of-the-art machine learning baselines with an accuracy and F1-score of up to 96\% and 89\% on the MediaEval benchmark \cite{boididou2015verifying} and the FakeNewsNet dataset \cite{shu2018fakenewsnet}, respectively.

The rest of this paper is organized as follows. In Section~\ref{sec:relatedworks}, we first review and discuss related works. In Section~\ref{sec:framework}, we detail MONITOR and especially feature extraction and selection. In Section~\ref{sec:experiments}, we present and comment on the experimental results that we achieve with respect to state-of-the-art methods. Finally, in Section~\ref{sec:conclusion}, we conclude this paper and outline future research.

\section{Related Works}
\label{sec:relatedworks}

%In this section, we review all non-image features and image features that are essential for checking the veracity of microblog posts.             

\subsection{Non-image Features}
Studies in the literature present a wide range of non-image features. These features may be divided into two subcategories, textual features and social context features. Textual features are extracted from the text content of messages, they are derived from the linguistics of a text to capture specific writing styles and the headlines that commonly occur in fake news content, such as lexical and syntactic features. 

To classify a message as fake or real, Castillo \textit{et al.} \cite{castillo2011information} capture prominent statistics in tweets, such as count of words, capitalized characters and punctuation, total number of words and characters. Beyond these features, lexical words expressing specific semantics or sentiments are also crucial clues to characterize the text, emotional marks (question marks and exclamation marks), and emoticons are also counted. Many sentimental lexical features are proposed in \cite{kwon2013prominent}, who utilize a sentiment tool called the Linguistic Inquiry and Word Count (LIWC) to count words in meaningful categories. 

Other works exploit syntactic features derived from the sentence level of rumors, such as the number of keywords, the sentiment score or polarity of the sentence. Features based on topic models are used to understand messages and their underlying relations within a corpus. Wu \textit{et al.} \cite{wu2015false} train a Latent Dirichlet Allocation model \cite{blei2003latent} with a defined set of topic features to summarize semantics for detecting rumors on the Sina Weibo microblogging platform.  

The social context reflects the interactions among different users and describes the propagating process of a rumor \cite{shu2018understanding}. Post content features represent the users’ social response in terms of stance. Social network features are extracted by constructing specific networks, such as diffusion \cite{kwon2013prominent} or co-occurrence networks \cite{ruchansky2017csi}.

Recent approaches detect fake news based on temporal-structure features. Kwon \textit{et al.} \cite{kwon2017rumor} studied the stability of features over time and found that, for rumor detection, linguistic and user features are suitable for early-stage, while structural and temporal features tend to have good performance in the long-term stage.
\subsection{Image Features}
Although images are widely shared on social networks, their potential for verifying the veracity of messages in microblogs is not sufficiently explored. Morris \textit{et al.} \cite{morris2012tweeting} assume that the user profile image has an important impact on information credibility published by this user. For images attached in messages, very basic features are proposed by \cite{wu2015false}, who define a feature called “has multimedia”  to mark whether the tweet has any picture, video or audio attached. Gupta \textit{et al.} \cite{gupta2013faking} propose a classification model to identify fake images on Twitter during Hurricane Sandy. However, their work is still based on textual content features. 

To automatically predict whether a tweet that shares multimedia content is fake or real, Boididou \textit{et al.} \cite{boididou2015verifying} propose the Verifying Multimedia Use (VMU) task. Textual and image forensics \cite{li2014segmentation} features are used as baseline features for this task. They conclude that Twitter media content is not amenable to image forensics and that forensics features do not lead to consistent VMU improvement \cite{boididou2018detection}. Finally, Jin \textit{et al.} \cite{jin2017multimodal} mostly focus on classification models for the problem rather than image features.

\section{MONITOR}

\label{sec:framework}

Microblog messages contain rich multimodal resources, such as text contents, surrounding social context, and attached image. Our focus is to leverage this multimodal information to determine whether a message is true or false. Based on this  idea, we propose a framework for verifying the veracity of messages. MONITOR's detailed description is presented in this section. 

\subsection{Multimodal Fusion Overview}

\begin{figure}[htbp]
	\centerline{\includegraphics[width=0.8\textwidth]{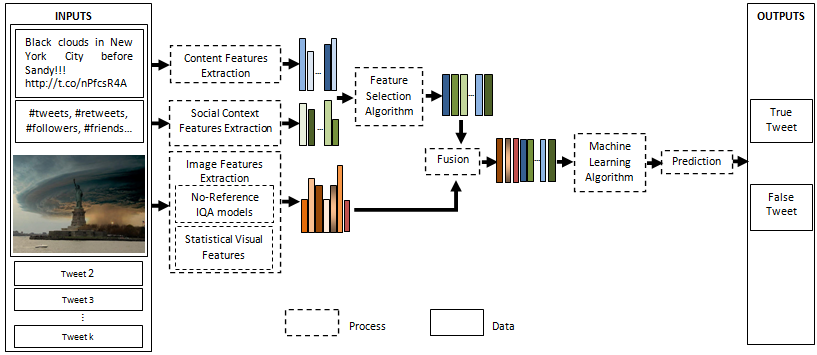}}
	\caption{Overview of MONITOR}
	\label{Fig2} 
\end{figure}

We define a message as a tuple of text, social context, and image content. MONITOR takes features from these modalities and aims to learn a multimodal fusion features vector as an aggregation of these aspects of the message. Figure.~\ref{Fig2} shows a general overview of MONITOR. 

It has two main stages:~1) Features extraction and selection. We extract several useful features from the message text and the social context, we then perform a feature selection algorithm to identify the relevant features, which form a first set of textual features. From the attached image, we drive statistics and efficient visual features inspired from the IQA field, which form a second set of image features;~2) Model learning. Textual and image features sets are then concatenated and normalized to form the fusion vector as the final multimodal representation of the message. Several machine learning classifiers may learn from the fusion vector to distinguish the veracity of the message (i.e., real or fake).

\subsection{Feature Extraction and Selection}

The feature extraction stage aims to represent message content and related auxiliary information in a formal measurable structure. To better choose features, we reviewed the best practices followed by information professionals (e.g., journalists) in verifying content generated by social network users. We based our thinking on relevant data from journalistic studies \cite{martin2014information} and the verification handbook \cite{silverman2014verification}. We define a set of features that are important to extract discriminating characteristics of rumors. These features are mainly derived from three principal aspects of news information: content, social context, and visual content of images.   

As for the feature selection process, it will only be applied to content and social context features sets to remove the irrelevant features that can negatively impact performance. Because our focus is the visual features set, we keep all these features in the learning process. 

\subsubsection{Message Content Features}  

Content features are extracted from the message's text. Aiming to arouse much attention and stimulate the public mood, rumor texts tend to have certain patterns in contrast to non-rumors. We extract characteristics such as the length of a tweet text and the number of its words. These characteristics also include statistics such as the number of exclamation and question marks, as well as binary features indicating the existence or not of emoticons. Furthermore, other features are extracted from the linguistics of a text, including the number of positive and negative sentiment words. For the English language, we use Liu and Hu's opinion lexicon list\footnote{https://www.cs.uic.edu/~liub/FBS/sentiment-analysis.html\#lexicon}, for German the Leipzig Affective Norms \cite{kanske2010leipzig}, and for Spanish the adaptation of ANEW \cite{redondo2007spanish}. Additional binary features indicate whether the text contains personal pronouns.

The veracity of the message text could also be related to its readability. We calculate a readability score between 1 and 100 using the Flesch Reading Ease method \cite{kincaid1975derivation}, the higher this score is, the easier the text is to read. For tweets written in a language for which the above features cannot be extracted, we consider the corresponding values to be missing.  Other features are extracted from the informative content provided by the specific communication style of the Twitter platform, such as the number of retweets, mentions(@), hashtags(\#), and URLs. 
\subsubsection{Social Context Features}  
The social context reflects the relationship between the different users and describes the process of spreading a rumor, therefore the characteristics of the social context are extracted from the behavior of the users and the propagation network. We capture several features from the users' profiles, such as number of followers and friends, number of tweets the user has authored, the number of tweets the user has liked, whether the user is verified by the social media, and whether the user has a profile image. 

We extract, also, features from the  propagation tree that can be built from tweets and re-tweets of a message, such as the depth of the re-tweet tree. Tables~\ref{Tab1} and \ref{Tab2}  depicts a description of a sets of content feature, and social context features extracted for each message.

    \begin{table}[!ht]
	\centering
	\begin{minipage}[t]{0.48\linewidth}\centering
		\caption{Content features}
		\label{Tab1}
		\begin{tabular}{|l|}
		\hline
		\thead{Description}\\
		\hline
		\# of characters, words\\
		\# of question mark (?), exclamation mark (!) \\
		\# of uppercase characters in the tweet text\\
		\# of positive, negative sentiment words \\
		\# of mentions(@username), hashtags(\#link), URLs\\
		\# of happy, sad mood emoticon \\
		\# first, second, third order pronoun\\
		The readability score of the tweet text\\
		\hline
		\end{tabular}
	\end{minipage}\hfill%
	\begin{minipage}[t]{0.48\linewidth}\centering
		\caption{Social context features}
		\label{Tab2}
		\begin{tabular}{|l|}
			\hline
		\thead{Description}\\
		\hline
		\# of followers, friends, posts the user has\\
		Friends/followers ratio, times listed the user has\\
		\# of re-tweets, likes that the tweet has obtained\\
		Whether the user shares a homepage URL\\
		Whether The user has their own profile image\\
		Whether the author has a verified account\\
		\# of Tweets the user has liked\\
		\hline
		\end{tabular}
	\end{minipage}
\end{table}

To improve the performance of MONITOR, we perform a feature selection algorithm on the features sets listed in Tables~\ref{Tab1} and \ref{Tab2}. The details of the feature selection process are discussed in Section~\ref{sec:experiments}.

\subsubsection{Image Features}  
\label{sec:imgfeat}

To differentiate between false and real images in messages, we propose to exploit visual content features and visual statistical features that are extracted from the joined images.

\paragraph{Visual Content Features.} Usually, a news consumer decides the image veracity based on his subjective perception, but how do we quantitatively represent the human perception of the quality of an image?. The quality of an image means the amount of visual degradations of all types present in an image, such as noise, blocking artifacts, blurring, fading, and so on. 

The IQA field aims to quantify human perception of image quality by providing an objective score of image degradations based on computational models \cite{maitre2017photon}. These degradations are introduced during different processing stages, such as image acquisition, process, compression, storage, transmission, decompression, display or even printing. Inspired by the potential relevance of IQA metrics for our context, we use these metrics in an original way for a purpose different from what they were created for. More precisely, we think that the quantitative evaluation of the quality of an image could be useful for veracity detection. 

IQA is mainly divided into two areas of research: first, full-reference evaluation; and second, no-reference evaluation. Full-reference algorithms compare the input image against a pristine reference image with no distortion. In no-reference algorithms, the only input is the image whose quality we want to measure, these algorithms compare statistical features of the input image against a set of features derived from an image database.

In our case, we do not have the original version of the posted image; therefore, the approach that is fitting for our context is the no-reference IQA metric. For this purpose, we use three no-reference algorithms that have been demonstrated to be highly efficient. 

%\begin{enumerate}

%\item 
The Blind/Referenceless Image Spatial Quality Evaluator (BRISQUE)  \cite{mittal2011blind} is trained on a database of images with known distortions, and is limited to evaluating the quality of images with the same type of distortion. BRISQUE is opinion-aware, which means that subjective quality scores accompany the training images.

%\item 
The Naturalness Image Quality Evaluator (NIQE) \cite{mittal2012making} is trained on a database of pristine images and can measure the quality of images with arbitrary distortion. NIQE is opinion-unaware and does not use subjective quality scores.

%\item 
The Perception based Image Quality Evaluator (PIQE) \cite{venkatanath2015blind}  is opinion-unaware and unsupervised (i.e., it does not require a trained model). PIQE can measure the quality of images with arbitrary distortion.

%\end{enumerate}  

\begin{figure}[htbp]%
	
	\centering
	\captionsetup[subfigure]{labelformat=empty,justification=centering}
	
	\begin{subfigure}[b]{0.18\textwidth}
	\includegraphics[width=\textwidth]{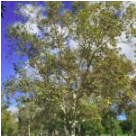}
	\caption{Original image\\ 13.7215}
	\label{}
	\end{subfigure}
	\quad
	\begin{subfigure}[b]{0.18\textwidth}
		\includegraphics[width=\textwidth]{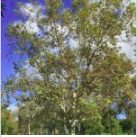}
		\caption{JPEG compression 22.6603}
		\label{}
	\end{subfigure}
	\quad
		\begin{subfigure}[b]{0.18\textwidth}
		\includegraphics[width=\textwidth]{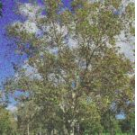}
		\caption{Gaussian Noise 28.5840}
		\label{}
	\end{subfigure}
	\quad
	\begin{subfigure}[b]{0.18\textwidth}
		\includegraphics[width=\textwidth]{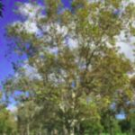}
		\caption{Median Blur \\4.1562}
		\label{}
	\end{subfigure}
	\caption{BRISQUE score  computed for a natural image and its distorted versions}%
	\label{Fig3}%
	\end{figure}

For example, Figure~\ref{Fig3} displays the  BRISQUE score computed for a natural image and its distorted versions (compression, noise and blurring distortions). The BRISQUE score is a  non-negative scalar in the range [1, 100]. Lower values of score reflect better perceptual quality of image.

\begin{figure}[htbp]%
	
	\centering
	\captionsetup[subfigure]{labelformat=empty,justification=centering}
	\begin{subfigure}[b]{0.16\textwidth}
		\centering
		\includegraphics[width=\textwidth]{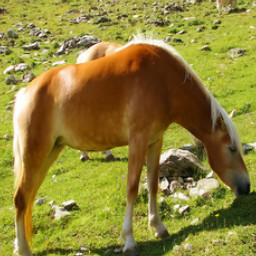}
		\caption{Real image 17.7778}
		\label{}
	\end{subfigure}
	\quad
	\begin{subfigure}[b]{0.16\textwidth}
		\centering
		\includegraphics[width=\textwidth]{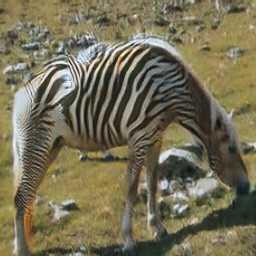}
		\caption{Fake image 22.0260}
		\label{}
	\end{subfigure}
	\quad\qquad
		\begin{subfigure}[b]{0.16\textwidth}
		\centering
		\includegraphics[width=\textwidth]{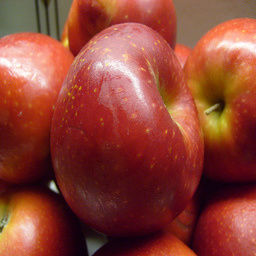}
			\caption{Real image 12.5000}
			\label{}
		\end{subfigure}
	\quad
		\begin{subfigure}[b]{0.16\textwidth}
		\centering
		\includegraphics[width=\textwidth]{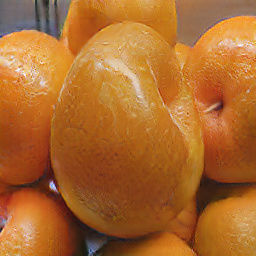}
			\caption{Fake image 22.5279}
			\label{}
		\end{subfigure}
	\caption{BRISQUE score  computed for real and fake generated GAN images}%
	\label{Fig4}%
\end{figure}

No-reference IQA metrics are not only used for traditional forgery detection, but are also good indicators for other types of image modifications, such as GAN-generated images. These techniques allow modifying the context and semantics of images in a very realistic way. Unlike many image analysis tasks, where both reference and reconstructed images are available, images generated by GANs may not have any reference image.  This is the main reason for using NR-IQA for evaluating this type of fake images, since these algorithms assess image quality without needing a reference nor its characteristics. Figure~\ref{Fig4} displays the BRISQUE score computed for real and fake images generated by image-to-image translation based on GANs \cite{CycleGAN2017}.

\paragraph{Statistical Features.} These are statistics of images attached to the text of a message. Similar to the statistical features of message content, some basic statistics of images proved to be distinctive in separating rumors and non-rumors. We define four statistical features from two aspects.

\textit{Number of Images:} In addition to the text, a user can post one, several, or no images. To denote this feature, we count the total number of images in a rumor event and the ratio of posts containing more then one image.

%	\item 
\textit{Spreading of Images:} During an event, some images are very replied and generate more comments than others. The ratio of such images is calculated to indicate this feature. Table~\ref{tab3} illustrates the description of proposed visual and statistical features to characterize the content of images. We use the whole set of these features in the learning process.   

\subsection{Model Training}
So far, we have obtained a first set of relevant textual features through a feature selection process performed on text content and social context features. We have also obtained a second set of image features composed of statistical and visual features driven from the field of image quality evaluation. These two sets of features are scaled, normalized, and concatenated to form the multimodal representation for a given message, which is fed to learn a supervised classifier for the rumor verification goal.
Several learning algorithms can be implemented for the classification task of message veracity. In the experimental part, we investigate the algorithms that provide the best performance. 

In summary, MONITOR takes as input training data with contents from three different modalities: text, social context and image. The output is the prediction label for each message instance to indicate it as true or false.
In the following section, we present empirical experiments to evaluate MONITOR's ability to assess message veracity.
\begin{table}[htbp]
	\caption{Description of image features}
	\begin{center}
		\begin{tabular}{|l|l|l|}
			\hline
			\thead{Type}&\thead{Feature}&\thead{Description}\\
			\hline
			\multirow{3}{*}{Visual}&BRISQUE&The BRISQUE score of a given image\\
			&PIQE&The PIQE score of a given image\\
			Features&NIQE&The NIQE score of a given image\\
			\hline
			\multirow{3}{*}{Statistical}&Count\_Img&The number of all images in a news event\\
			&Ratio\_Img1&The ratio of the multi-image tweets in all tweets\\
			Features&Ratio\_Img2&The ratio of image number to tweet number\\
			&Ratio\_Img3&The ratio of the most widespread image in all distinct images\\
			\hline
		\end{tabular}
		\label{tab3}
	\end{center}
\end{table}
\section{Experiments}
\label{sec:experiments}
In this section, we conduct extensive experiments on two  public datasets, to validate the effectiveness of the image features derived from IQA (Section~\ref{sec:imgfeat}) and the relevance of fusing several features.% for news verification task on microblogs. 
First, we present statistics about the datasets we used. Then, we describe the experimental settings: a brief review of state-of-the-art features for news verification and a selection of the best of these textual features as baselines. Finally, we present experimental results and analyze the features to achieve insights into MONITOR.
\subsection{Datasets}
To evaluate MONITOR's performance, we conduct experiments on two well-established public benchmark datasets for rumor detection, i.e., the MediaEval Verifying Multimedia Use \cite{boididou2015verifying} and the FakeNewsNet   \cite{shu2018fakenewsnet} benchmarks. Their statistics are shown in Table~\ref{tab4}.

\subsubsection{MediaEval}
%We actually only use the dataset from the  , not the whole benchmar. This dataset 
 is collected from Twitter and includes all three characteristics: text, social context and images . It is designed for message-level verification. The dataset has two parts: a development set containing about 9,000 rumor and 6,000 non-rumor tweets from 17 rumor-related events; a test set containing about 2,000 tweets from another batch of 35 rumor-related events. We remove tweets without any text or image, thus obtaining a final dataset including 411 distinct images associated with 6,225 real and 7,558 fake tweets, respectively.
 
\subsubsection{FakeNewsNet}
 is one of the most comprehensive fake news detection benchmark. Fake and real news articles are collected from the fact-checking websites PolitiFact and GossipCop. Since we are particularly interested in images in this work, we extract and exploit the image information of all tweets. To keep the dataset balanced, we randomly choose 2,566 real and 2,587 fake news events. After removing tweets without images, we obtain 56,369 tweets and 59,838 images.
 
% The statistical details of both dataset are shown in .
\begin{table}
	\caption{MediaEval and FakeNewsNet statistics}\label{tab4}
	\centering
	\begin{tabular}{|c|c|c|c|c|}
		\hline
		\multirow{2}{*}{Dataset}&\multirow{2}{*}{Set}&\multicolumn{2}{c|}{Tweets}&\multirow{2}{*}{Images}\\
		\cline{3-4}
		&&Real&Fake&\\
		\hline \hline
		MediaEval&Training Set&5,008&6,841&361\\
		&Testing Set&1,217&717&50\\
		\hline
		FakeNewsNet&Training Set&25,673&19,422&47,870\\
		&Testing Set&6,466&4,808&11,968\\	
		\hline
	\end{tabular}
\end{table}

\subsection{Experimental Settings}

%\begin{enumerate}[label=\Alph*)]  Une section ne peut pas être juste une liste

\subsubsection{Baseline Features} We compare the effectiveness of our feature set with the best textual features from the literature. First, we adopt the 15 best features extracted by Castillo \textit{et al.} from aspects of message content, user, topic and propagation tree, to analyze the information credibility of news propagated through Twitter \cite{castillo2011information}.
%T, Castillo \textit{et al.} propose 15 best features extracted from aspects of message content, user, topic, and propagation tree \cite{castillo2011information}. 
We also collect a total of 40 additional textual features proposed in the literature \cite{gupta2013faking,gupta2012evaluating,kwon2013prominent,wu2015false}, which are extracted from text content, user information and propagation properties (Table~\ref{Tab5}).

\subsubsection{Feature Sets} 

The features labeled \textit{Textual} are the best features selected among message content and social context features (Tables~\ref{Tab1} and \ref{Tab2}). We select them  with the information gain ratio method~\cite{karegowda2010comparative}, which is commonly used for measuring the goodness of attributes in decision tree learning, over both datasets. It helps select a subset of 15 relevant textual features with an information gain larger than zero (Table~\ref{Tab6}). 
 
The features labeled \textit{Image} are all the image features listed in Table~\ref{tab3}. The features labeled \textit{MONITOR} are the feature set that we propose, consisting of the fusion of textual and image feature sets. The features labeled \textit{Castillo} are the above-mentioned best 15 textual features. %  proposed in \cite{castillo2011information}. 
Eventually, the features labeled \textit{Wu} are the 40 textual features identified in literature. % \cite{kwon2013prominent,gupta2013faking,gupta2012evaluating,wu2015false}.

%\end{itemize}
 \begin{table}[!ht]
	\centering
	\begin{minipage}[t]{0.48\linewidth}\centering
		\caption{40 features from the literature}
		\label{Tab5}
		\begin{tabular}{|l|}
			\hline
			\thead{Feature}\\
			\hline
			Fraction of Question, Exclamation Mark,\\
			Count of Message,\\
			Average Word, Character Length,\\
			Fraction of First, Second, Third Pronouns,\\
			Fraction of URL,@,\#,\\
			Count of Distinct URL,@,\#,\\
			Fraction of Popular URL,@, \#,\\
			Whether the Tweet includes pictures,\\
			Average Sentiment Score,\\
			Fraction of Positive, Negative Tweets,\\
			Count of Distinct People, Location, Organization,\\
			Fraction of People, Location, Organization,\\
			Fraction of Popular People, Location, Organization.\\
			Count of Distinct Users, Fraction of Popular Users,\\
			Count Followers, Followees, Posted Tweets,\\
			Whether the User Has Facebook Link,\\
			Fraction of Verified User, Organization,\\
			Count comments on the original message\\
			Time interval between original message and repost\\
			\hline
		\end{tabular}
	\end{minipage}\hfill%
	\begin{minipage}[t]{0.48\linewidth}\centering
		\caption{Best textual features selected %using a feature selection process 
			by gain ratio}
		\label{Tab6}
		\begin{tabular}{|l|l|}
				\hline
			\thead{MediaEval}&\thead{FakeNewsNet}\\
			\hline
			Tweet\_Length&Tweet\_Length \\
			Num\_Negsentiwords&Num\_Words\\
			Num\_Mentions&Num\_Questmark\\
			Num\_URLs&Num\_Uppercasechars\\
			Num\_Words&Num\_Exclammark\\
			Num\_Uppercasechars&Num\_Hashtags\\
			Num\_Hashtags&Num\_Negsentiwords\\
			Num\_Exclammark&Num\_Possentiwords\\
			Num\_Thirdorderpron&Num\_Followers\\
			Times\_Listed&Num\_Friends\\
			Num\_Tweets&Num\_Favorites\\
			Num\_Friends&Times\_Listed\\
			Num\_Retweets&Num\_Likes\\
			Has\_Url&Num\_Retweets\\
			Num\_Followers&Num\_Tweets\\
			\hline
		\end{tabular}
	\end{minipage}
\end{table}

\subsubsection{Classification Model} To assess the robustness of our proposal, we execute various learning algorithms for each feature set. The best results are achieved by four supervised classification models: decision trees, KNNs, SVMs and  random forests. We use their Scikit-learn   library  for Python~\cite{scikit-learn} implementation. Training and validation is performed for each model through a 5-fold cross validation. 
Note that, for MediaEval, we retain the same data split scheme. For FakeNewsNet, we randomly divide data into training and testing subsets with the ratio 0.8:0.2. Table~\ref{tab7} present the results of our experiments. We use \textit{Accuracy}, \textit{Precision}, \textit{recall} and  $F_{1}$ score to evaluate the overall prediction performance.

\begin{table}[h]
	\caption{Classification results}\label{tab7}
	%\resizebox{\linewidth}{!}{
	\centering
	\begin{tabular}{|c|c|c|c|c|c||c|c|c|c|}
		\cline{1-10}
		&&\multicolumn{4}{c||}{\textbf{MediaEval}}&\multicolumn{4}{c|}{\textbf{FakeNewsNet}}\\
		\cline{3-10}
		\multirow{2}{*}{\textbf{Model}} & \multirow{2}{*}{\textbf{Feature sets}} & &&&&&&&\\
		%\cline{4-6} \cline{8-10}
		&  &\textbf{Accuracy} & \textbf{Precision}&\textbf{Recall}&  $\bm{F_{1}}$ &\textbf{Accuracy}	&\textbf{Precision}&\textbf{Recall}&  $\bm{F_{1}}$ \\
		\hline
		\multirow{5}{*}{Decision} 
		& Textual&0.673&0.672&0.771&0.718&0.699&\textbf{0.647}&0.652&0.65\\
		& Image&0.632&0.701&0.639&0.668&0.647&0.595&0.533&0.563\\
		&MONITOR&\textbf{0.746}&\textbf{0.715}&\textbf{0.897}&\textbf{0.796}&\textbf{0.704}&0.623&\textbf{0.716}&\textbf{0.667}\\ \cline{2-10} 
		
		Trees&Castillo&0.643&0.711&0.648&0.678&0.683&0.674&0.491&0.569\\	 
		&Wu&0.65&0.709&0.715&0.711&0.694&0.663&0.593&0.627\\
		
		\hline
		\multirow{5}{*}{KNN}
		& Textual&0.707&0.704&0.777&0.739&0.698&0.67&0.599&0.633\\
		& Image&0.608&0.607&0.734&0.665&0.647&0.595&0.533&0.563\\
		&MONITOR&\textbf{0.791}&\textbf{0.792}&\textbf{0.843}&\textbf{0.817}&\textbf{0.758}&\textbf{0.734}&\textbf{0.746}&\textbf{0.740}\\ \cline{2-10} 
		& Castillo&0.652&0.698&0.665&0.681&0.681&0.651&0.566&0.606\\	 
		&Wu&0.668&0.71&0.678&0.693&0.694&0.663&0.593&0.627\\
		\hline
		\multirow{5}{*}{SVM}
		& Textual&0.74&0.729&0.834&0.779&0.658&\textbf{0.657}&0.44&0.528\\
		& Image&0.693&0.69&0.775&0.73&0.595&0.618&0.125&0.208\\
		&MONITOR&\textbf{0.794}&\textbf{0.767}&\textbf{0.881}&\textbf{0.82}&\textbf{0.704}&0.623&\textbf{0.716}&\textbf{0.667}\\ \cline{2-10}
		& Castillo &0.702&0.761&0.716&0.737&0.629&\textbf{0.687}&0.259&0.377\\	 
		&Wu&0.725&0.763&0.73&0.746&0.642&0.625&0.394&0.484\\
		\hline
		\multirow{5}{*}{Random}
		& Textual&0.747&0.717&0.879&0.789&0.778&0.726&0.768&0.747\\
		& Image&0.652&0.646&0.771&0.703&0.652&0.646&0.771&0.703\\
		&MONITOR&\textbf{0.962}&\textbf{0.965}&\textbf{0.966}&\textbf{0.965}&\textbf{0.889}&\textbf{0.914}&\textbf{0.864}&\textbf{0.889}\\ \cline{2-10} 
		Forest& Castillo&0.702&0.727&0.723&0.725&0.714&0.669&0.67&0.67\\	 
		&Wu&0.728&0.752&0.748&0.75&0.736&0.699&0.682&0.691\\
		\hline
	\end{tabular}
	
\end{table}

\subsection{Classification Results}
From the classification results recorded in Tables~\ref{tab7}, we can make the following observations.
\subsubsection{Performance Comparison} With MONITOR, using both image and textual feature allows all classification algorithms to achieve better performance than baselines. Among the four classification models, the random forest generates the best accuracy: 96.2\% on MediaEval and 88.9\% on FakeNewsNet. They indeed perform 26\% and 18\% better than Castillo and 24\% and 15\%  than Wu,  %\cite{kwon2013prominent,gupta2013faking,gupta2012evaluating,wu2015false}, 
still on MediaEval and FakeNewsNet, respectively. 

Compared to the 15 ``best'' textual feature set, the random forest improves the accuracy by more than 22\% and 10\% with image features only. Similarly, the other three algorithms achieve %with image features 
an accuracy gain between 5\% and 9\% on MediaEval and between 5\% and 6\% on FakeNewsNet.
Compared to the 40 additional textual features, all classification algorithms generate a lower accuracy when using image features only. This is due to the lack of social context and also because textual features are selected from a wide range of textual properties.

While image features play a crucial role in rumor verification, we must not ignore the effectiveness of textual features. The role of image and textual features is complementary. When the two sets of features are combined, performance is significantly boosted.   

\subsubsection{Illustration by Example} To more clearly show this complementarity, we compare the results reported by MONITOR and single modality approaches (textual and image). The fake rumor messages from Figure~\ref{Fig1} are correctly detected as false by MONITOR, while using either only textual or only image modalities yields a true result.

In the tweet from Figure~\ref{Fig1}(a), the text content solely describes the attached image without giving any signs about the veracity of the tweet. This is how the textual modality identified this tweet as real. It is the attached image that looks quite suspicious.% and could most probably be doctored. 
By merging the textual and image contents, MONITOR can identify the veracity of the tweet with a high score, exploiting some clues from the image to get the right classification.

The tweet from Figure~\ref{Fig1}(b) is an example of a rumor correctly classified by MONITOR, but incorrectly classified when only using the visual modality. The image seems normal and the complex semantic content of the image is very difficult to capture by the image modality. However, the words with strong emotions in the text indicate that it might be a suspicious message. By combining the textual and image modalities, MONITOR can classify the tweet with a high confidence score. This tweet presents a particular type of rumor that is very challenging to identify, because the attached image has been misused from its original context: the two boys were actually photographed in Vietnam in 2007 and have nothing to do with the Nepal Earthquake in 2015.

\subsection{Feature Analysis}
The advantage of our approach is that we can achieve some elements of interpretability, especially to explain the contribution of image and textual features in the prediction process. Thus, we conduct an analysis to illustrate the importance of each feature set. 
To identify what features have the most predictive power, we depict the first most 15 important features achieved by the random forest. Variables of high importance are drivers of the classification and their values have a significant impact on the prediction.

\begin{figure}%
		\centering
	\begin{subfigure}[b]{0.49\textwidth}
		\centering
		\includegraphics[width=0.50\textwidth]{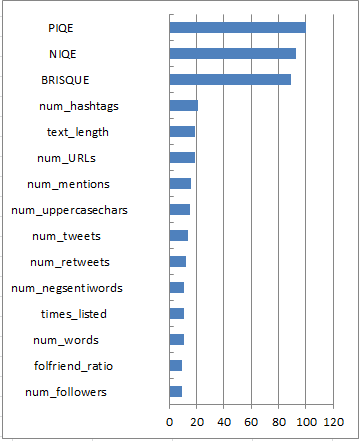}
		\caption{MediaEval}
		\label{}
	\end{subfigure}
	%\quad
		\begin{subfigure}[b]{0.49\textwidth}
		\centering
		\includegraphics[width=0.45\textwidth]{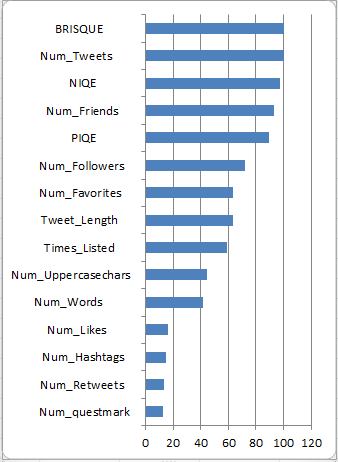}
		\caption{FakeNewsNet}
		\label{}
	\end{subfigure}
	\caption{Top-15 important variables}%
	\label{Fig5}%
\end{figure}

Figure~\ref{Fig5} shows that, for both datasets, visual characteristics are in the top five features. The remaining features are a mix of text content and social context features. These results validate the effectiveness of the IQA image features issued, as well as the the importance of fusing several modalities in the process of rumor verification.

\begin{figure}%
	\centering
	\begin{subfigure}[b]{0.48\textwidth}
		\centering
		\includegraphics[width=0.88\textwidth]{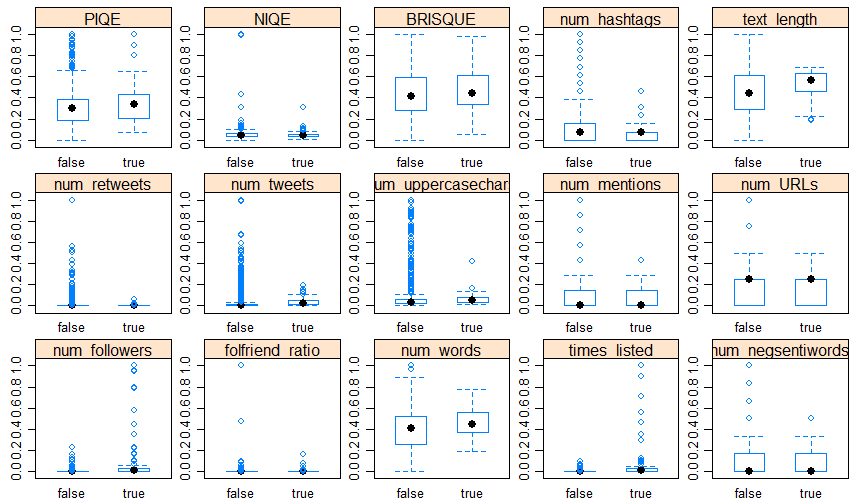}
		\caption{MediaEval}
		\label{}
	\end{subfigure}
	%\quad
	\begin{subfigure}[b]{0.48\textwidth}
		\centering
		\includegraphics[width=0.83\textwidth]{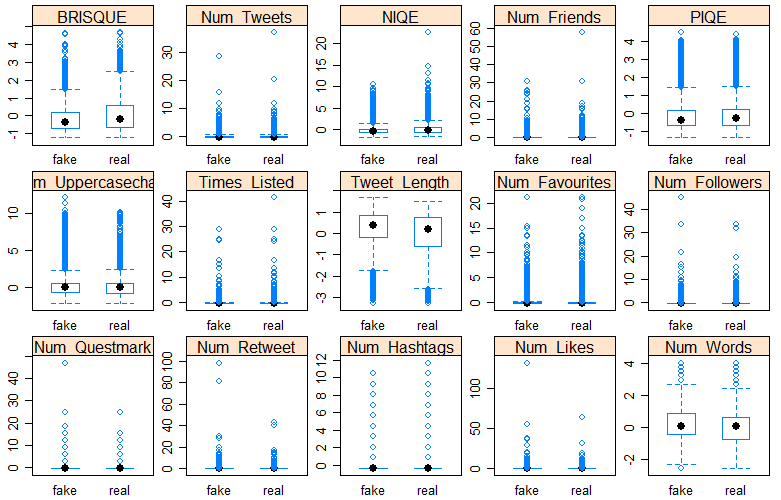}
		\caption{FakeNewsNet}
		\label{}
	\end{subfigure}
	\caption{Distribution of true and false classes for top-15 important features}%
	\label{Fig6}%
\end{figure}

To illustrate the discriminating capacity of these features, we deploy box plots for each of the 15 top variables on both datasets. Figure~\ref{Fig6} shows that several features exhibit a significant difference between the fake and real classes, which explains our good results.% This explains our results in tweets veracity assessment.
\section{Conclusion and Perspectives}
\label{sec:conclusion}
To assess the veracity of messages posted on social networks, most machine learning %feature-based
techniques ignore the images attached to messages. In this paper, to improve the performance of the message verification, we propose a multimodal fusion framework called MONITOR that uses features extracted from the textual content of the message, the social context, and also image features have not been considered until now. Extensive experiments conducted on the MediaEval benchmark and FakeNewsNet dataset demonstrated that: 1) the image features that we introduce play a key role in message veracity assessment; and 2) no single homogeneous feature set can generate the best results alone. They also show that with a classification accuracy higher than 96\% on MediaEval, and 89\% on FakeNewsNet, MONITOR outperforms state-of-the-art machine learning methods.

Our future research includes two directions. First, we currently fuse textual, context, and image features into a single vector, which is called early fusion. By combining classifiers instead, we also plan to investigate so-called late fusion.
Second, deep learning models are capable of learning from representations of both text and images. In particular, recurrent neural networks (RNNs) are widely used in sentence representation and convolutional neural networks (CNNs) are efficient for image representation. Combining RNNs and CNNs  could thus be useful for detecting rumors. However, we would like to compare their performance with MONITOR's to study the tradeoff between classification accuracy, computing complexity, and explainability.

%BibTeX users should specify bibliography style 'splncs04'.
%\bibliographystyle{splncs04}
%\bibliography{biblio}
%

\end{document}